\documentclass[AMA,Times2COL]{WileyNJDv5}
\startpage{1}
\usepackage{dblfloatfix}
\usepackage{placeins}
\usepackage{afterpage}
\usepackage{fancyhdr}
\begin{document}
\title{Transport Layer-Driven Optimization of Color and Efficiency in TADF OLEDs for Large-Area Applications}
\author[1]{Manish Kumar$^{*}$}
\author[2]{Jawid Nikan}
\author[3]{Luiz Pereira$^{*}$}
\address[1]{\orgdiv{Wihuri Physical Laboratory, Department of Physics and Astronomy}, \orgname{University of Turku}, \orgaddress{\state{Turku}, \country{Finland-20014}}}
\address[2]{\orgdiv{Department of Molecular Electronics}, \orgname{Max-Planck Institute for Polymer research}, \orgaddress{\state{Mainz}, \country{Germany}}}
\address[3]{\orgdiv{Department of Physics}, \orgname{University of Aveiro}, \orgaddress{\state{Aveiro}, \country{Portugal}}}
\corres{Corresponding author Manish Kumar. Luiz Pereira \email{manish.kumar@utu.fi, Luiz@ua.pt}}
\abstract[Abstract]{Balancing structural simplicity with high efficiency is crucial for practical applications of organic light-emitting diodes (OLEDs), especially in large-area lighting devices. Optimizing the charge-transport layers in multilayer OLEDs controls the recombination zone within the emissive layer, thereby enhancing the device's performance. Herein, we experimentally and computationally investigate how the thicknesses of the electron and hole transport layers affect the performance of orange-red thermally activated delayed fluorescence (TADF) OLEDs. Using a TADF emitter, 2-[4-(diphenylamino)phenyl]-10,10-dioxide-9H-thioxanthen-9-one (TXO-TPA), at 10 wt\% concentration in  4,4-Bis(N-carbazolyl)-1,1-biphenyl (CBP) host, we achieve a maximum external quantum efficiency ($\eta_{EQE}$) of 20.86\% and a peak luminance of 32,800 cd m$^{-2}$. Using a combination of drift-diffusion and optical simulations, we reveal how variations in hole- and electron-transport-layer thicknesses modulate the charge-carrier balance and shift the recombination zone within the emissive layer. This tuning not only optimizes exciton formation but also enables color tuning from green to red without altering the emissive material. Our findings highlight the critical role of transport layer engineering in achieving efficient, color-controlled OLEDs with simplified architectures. Furthermore, we successfully demonstrate a large-area OLED device (~20 cm$^{2}$), achieving an impressive $\eta_{EQE}$ of >17\%, underscoring the scalability and practicality of our approach for general lighting applications.}

\maketitle
\pagestyle{empty}
\thispagestyle{empty}

\section{Introduction}\label{sec1}
Organic Light-Emitting Diodes (OLEDs) continue to advance significantly due to their superior brightness, ease of fabrication, thinness, and high contrast, making them ideal for various lighting and display applications\cite{miao2024microdisplays, hong2021brief}. Current OLED lighting applications face critical challenges, including complex device architectures, efficiency roll-off at high luminance, and limited operational stability, hindering their large-scale commercialization. Recent research has increasingly focused on simplifying OLED structures by employing high-performance thermally activated delayed fluorescence (TADF) emitters to reach nearly 100\% internal quantum efficiency ($\eta_{IQE}$) through efficient triplet harvesting via reverse intersystem crossing (RISC)\cite{uoyama2012highly}. TADF-based OLEDs achieved more than 40\% external quantum efficiency ($\eta_{EQE}$) due to triplet harvesting through the reverse intersystem crossing (RISC) process\cite{fu2023realizing, park2023high, meng2023high, fan2023ultrapure, cheng2024efficient, liu2018all}. However, current TADF-based OLEDs typically have complex device architectures with multiple organic layers and suffer from efficiency roll-off\cite{kumar2019effect, van2021origin, murawski2013efficiency}, restricting their practical applications, especially in large-area lighting. Consequently, a balance between structural simplicity and high efficiency is necessary. Wetzelaer and colleagues have continued work on simplifying the device structure via single-emissive-layer devices and charge balance by optimizing electrodes and keeping the organic stack <100 nm\cite{kotadiya2019efficient, tan2026highly}.

Optimizing the thicknesses of the transport layers (hole transport layer: HTL and electron transport layer: ETL) is central to enhancing OLED performance. These layers critically influence the formation of the recombination zone, which results in device performance and lifetime. Moreover, simulation techniques, such as drift-diffusion modeling, have emerged as powerful tools to elucidate the intricate interplay between layer thicknesses and device characteristics. In a device, carrier (hole and electron) mobility and their balance are considered the most critical factors for performance because they influence the practical figures of merit, such as the driving voltage, efficiency, and lifetime\cite{rhee2014effect, culligan2006effect, lee2008influence, forrest2020organic, mehes2020physics, anthopoulos2003highly, kotadiya2019efficient}. In organic materials, the hole mobility is very high compared to electron mobility, leading to a hole accumulation near the EML-ETL interface without exciton generation, resulting in device degradation\cite{rhee2014effect, culligan2006effect, lee2008influence}. Earlier, we demonstrated that the thickness and electrical carrier mobility of both HTL and ETL govern the charge transport in the emissive layer and recombination for efficient solution-processed TADF OLEDs\cite{kumar2020towards}, which, in consequence, modulates the recombination region. Confining all the excitons and balanced charge carriers is essential to extract better performance from EML\cite{kumar2020towards}. 

The aforementioned are primarily applicable to non-doped emitter-based devices. To prevent concentration quenching in state-of-the-art TADF-based OLEDs, the EML is a matrix of TADF emitter doped in a wide-energy-gap host material that should be able to provide charge carriers to the emitter and confine the exciton\cite{holmes2003blue, jankus2014highly}. In this host-guest framework, this energy transfer, when optimized, can lead to a theoretical maximum IQE of 100\% \cite{wu2021role, dos2016using, stavrou2024key, dias2013triplet}. Upon doping, the host-guest matrix's charge mobilities change due to interactions between materials with different electrical properties \cite{cheng2019carrier, cui2017long, kumar2020mixed, liu2007charge, kumar2020insights}. When optimized, charge transport in EML can enhance exciton recombination and improve efficiency \cite{jou2018investigation}. However, in the usual host-guest system, the recombination zone is located in the EML near the ETL. This reduces the width of the recombination zone and total recombination probability. This can be avoided by employing HTL and ETL carrier mobilities that are nearly equivalent to enhance balanced hole and electron transport. In addition, considering the final carrier mobility in the EML, the HTL and ETL thicknesses can be modulated to improve recombination-region effects.

While red TADF OLEDs have achieved 40\% EQE with recent high-purity emitters, these often involve elaborate device structures or novel materials\cite{pu2025sulfur}. Here, we aim for a simpler device with competitive efficiency. In this work, we demonstrate a simplified yet efficient TADF OLED structure employing the orange-red emitter TXO-TPA doped into a host CBP. Through systematic experimental investigation and simulation studies, we optimize HTL and ETL thicknesses to enhance carrier balance, exciton recombination, and device performance. Our three-layer OLED achieves a peak luminance of 32,800 cd m$^{-2}$ and $\eta_{EQE}$ of 20.86\%, for such simplified red-orange TADF OLED devices. Furthermore, we successfully demonstrate efficient large-area OLED fabrication ($\sim$20 cm$^{2}$), underscoring the potential of transport layer engineering in practical lighting applications. This simplified device structure not only demonstrates superior efficiency but also facilitates the large-scale manufacturing required for commercial general lighting applications. Thus, our study highlights the viability of three-layer, TADF-based OLED architectures as cost-effective and sustainable solutions for high-quality indoor and outdoor lighting.

\section{Results and Discussion}\label{sec2}
\begin{figure}[htpb]
\centerline{\includegraphics[scale=0.3]{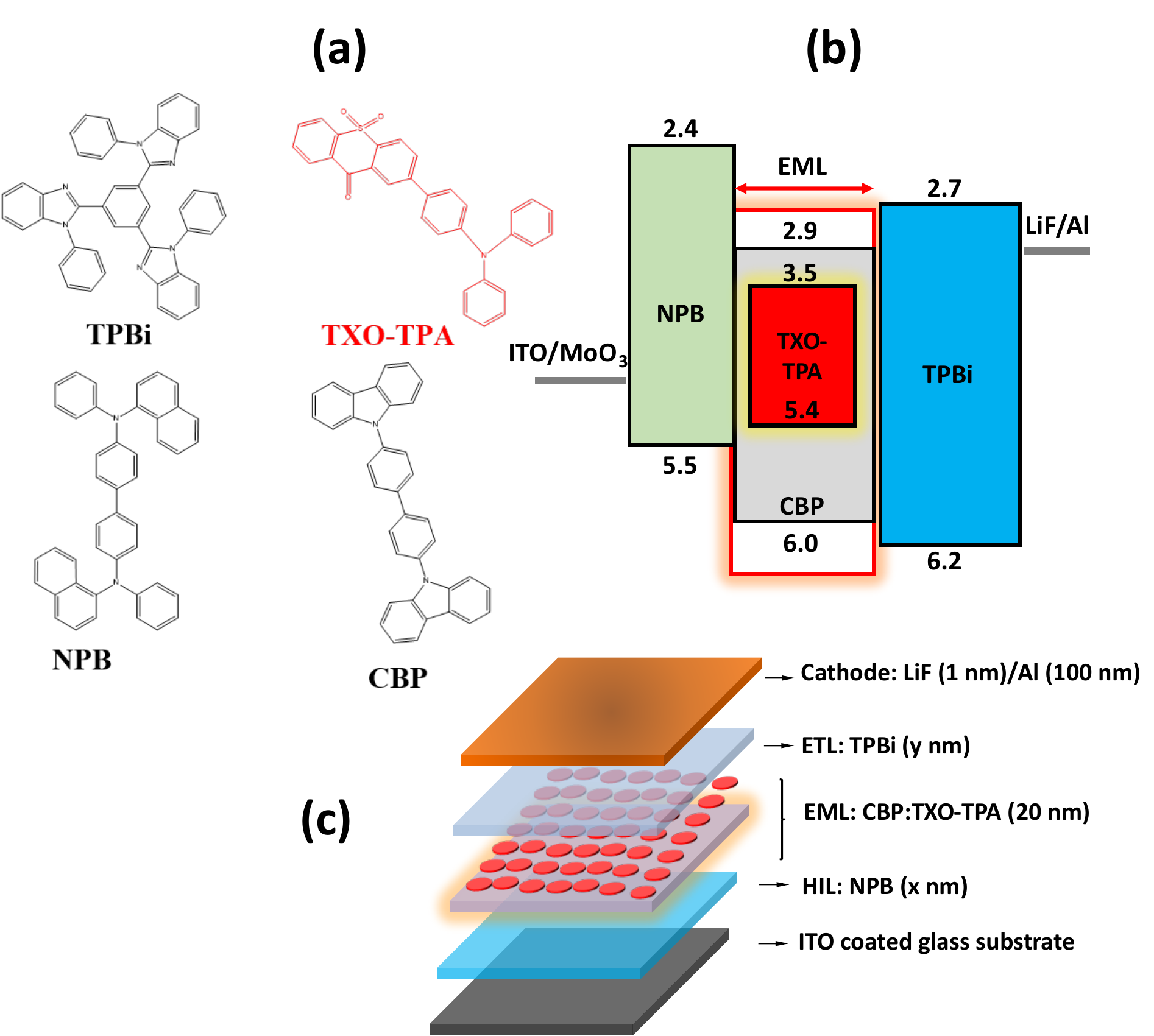}}
\caption{(a) Chemical structure of the materials used in OLED fabrication, (b) related energy levels (HOMO and LUMO) for ITO/MoO$_3$, NPB (N,N\ensuremath{{}^{\prime}}-Di(1-naphthyl)-N,N\ensuremath{{}^{\prime}}-diphenyl-(1,1\ensuremath{{}^{\prime}}-biphenyl)-4,4\ensuremath{{}^{\prime}}-diamine), CBP (4,4\ensuremath{{}^{\prime}}-Bis(9-carbazolyl)-1,1\ensuremath{{}^{\prime}}-biphenyl, 4,4-N,N\ensuremath{{}^{\prime}}-Dicarbazole-1,1\ensuremath{{}^{\prime}}-biphenyl), TXO-TPA (2-[4-(diphenylamino)phenyl]-10,10-dioxide-9H-thioxanthen-9-one), TPBi (2,2',2"-(1,3,5-Benzinetriyl)-tris(1-phenyl-1-H-benzimidazole)), and LiF/Al. The device structure used in this work (from bottom to top): substrate/ITO (100 nm)/ MoO$_{3}$ (5 nm)/ NPB ($d_{HTL}$ nm)/ CBP: TXO-TPA ($d_{EML}$ nm)/ TPBi ($d_{ETL}$ nm)/ LiF (1 nm)/ Al (100 nm). Here, MoO$_{3}$ was used as the hole injection layer (HIL), NPB as the hole transport layer (HTL), 10wt\% TXO-TPA in CBP as the emissive layer (EML), TPBi as the electron transport layer (ETL), and LiF as the electron injection layer (EIL) and Al as the bottom electrode. The illuminated red box illustrates the recombination zone that shifts towards the center of the EML from the ETL/EML interface, increasing the efficiency in different devices, and (c) the device architecture. 
\label{fig1}}
\end{figure}

OLEDs were fabricated in a multilayer structure employing the host-guest concept for the EML with an orange-red TXO-TPA TADF dopant in the CBP. TXO-TPA was chosen as the previously reported red-orange dopant\cite{wang2014novel}. The device structure herein is ITO/MoO$_{3}$ (5 nm)/NPB (x nm)/ EML (20 nm)/ TPBi (y nm)/LiF (1 nm)/Al (100 nm), where x = 20 nm or 30 nm and y= 20 nm or 30 nm (Figure~\ref{fig1}). The emitter concentration (10 wt\%) was optimized to avoid concentration quenching.  

\subsection{Theoretical Simulation}\label{sec2.1}
The device simulations assumed that the injection/transport of electrical carriers into the device is governed by the typical drift-diffusion model. The model considers surface injection/recombination current in organic semiconducting materials. Scott and Malliaras\cite{scott1999charge} propose a net injection current, given by:

\begin{equation}
J = 4 \Phi^2 N_0 q \mu E \exp\left(-\frac{\phi_b}{kT}\right) \exp\left(\sqrt{f}\right)
\label{eq:current_density}
\end{equation}
with
\[
\Phi = \frac{1}{f} + \frac{1}{\sqrt{f}} + \frac{\sqrt{1 + 2\sqrt{f}}}{f}
\]
and
\[
f = \frac{q^3 E}{4 \pi \varepsilon kT}
\]
where \(f\) is the reduced electrical field, \(\phi_b\) is the Schottky barrier height, and \(N_0\) is the density of the active electrical energy levels in the organic material.
The electrical carrier mobility for organic semiconductors follows the well-known Poole-Frenkel mobility, which is field and temperature-dependent. Considering only the field dependence, the mobility can be expressed as:
\[
\mu = \mu_0 \exp\left(\beta \sqrt{E}\right)
\]
where \(\beta\) includes the temperature \(T\) (assumed constant for all devices) and the Boltzmann constant \(k_B\).

The usual way to solve the theoretical model for electrical carrier injection, transport, and recombination is to treat the device as a set of "slices" from anode to cathode. Additionally, we assumed a one-dimensional Poisson equation to describe the evolution of the electrical field across the device, i.e.,
\[
\frac{\partial E}{\partial x} = \frac{q (p - n)}{\varepsilon}
\]
and
\[
\int_0^d E\,dx = V - V_{bi},
\]
where \(p\) and \(n\) are the density of holes and electrons, and \(V_{bi}\) is the built-in voltage. Under these conditions, and considering that electrical carrier transport occurs through both drift (electric field-driven) and diffusion (density gradient-driven), the current density due to both mechanisms is given by:

\begin{equation}
J_{n,p} = q \left(\mu_{n,p} E \cdot (n,p) + D_{n,p} \frac{\partial (n,p)}{\partial x}\right)
\label{eq:drift_diffusion}
\end{equation}
where \(J_{n,p}\) is the current density for electrons (\(n\)) or holes (\(p\)), \(q\) is the elementary charge, \(\mu_{n,p}\) denotes the mobility of electrons or holes, \(E\) is the electric field, \(n\) and \(p\) are the electron and hole densities, \(D_{n,p}\) is the diffusion coefficient for electrons or holes, and \(\frac{\partial (n,p)}{\partial x}\) represents the density gradient (diffusion term).

Formally, electrons and holes recombine following Langevin's theory with a recombination rate R, which can be given by\cite{jaffe1940theory, zhang2014tenfold}:
\begin{equation}
R = \frac{q}{\varepsilon} \left(\mu_p + \mu_n\right) n p
\label{eq:recombination}
\end{equation}
where \(R\) is the recombination rate, \(q\) is the elementary charge, \(\varepsilon\) is the permittivity, \(\mu_p\) and \(\mu_n\) are the hole and electron mobilities, and \(n\) and \(p\) are the electron and hole densities.
Inside a single organic layer, we can consider that the variation of the electrical carrier density with time can be expressed by the net contribution from drift, diffusion, and recombination, according to the continuity equation:
\[
\frac{\partial (n,p)}{\partial t} = \frac{1}{q} \frac{\partial J_{n,p}}{\partial x} - R
\]
where \(n\) and \(p\) are the electron and hole densities, \(J_{n,p}\) is the electron or hole current density, \(q\) is the elementary charge, and \(R\) is the recombination rate. The density of carriers in the device modulates the electric field, which in turn, changes the carrier distribution. Therefore, we need to start by considering all processes that determine the electric field variations and electron/hole densities at each position over time. When considering the complete device architecture in a multilayer structure, electrons and holes can face an energy barrier at the interface between two adjacent layers (primarily differences in HOMO / LUMO levels). The hole current density at the heterojunction interface can be written as\cite{wang2008current}:
\[
J_{p, \text{cross}} = q p_h E_{\text{int}} \sqrt{\mu_{p,h} \mu_{p,e}} \times \Gamma(\Delta E)
\]
where \(\Delta E = E_{\text{HOMO},h} - E_{\text{HOMO},e} - E_{\text{int}} \times d_m\), and \(\Gamma(\Delta E)\) is a function depending on \(\exp\left(-\frac{x_h^2}{2 \sigma_h}\right)\). A similar model can be derived for electrons. Then, the exciton density \(\chi\) diffusion, decay, and recombination can be expressed by:

\begin{equation}
\frac{\partial \chi}{\partial t} = R(x) \times \Gamma_{\text{spin}} - \frac{\chi}{\tau} + D_\chi \frac{\partial^2 \chi}{\partial x^2} - K_{\text{quench}}(x) \times \chi
\label{eq:exciton_dynamics}
\end{equation}
With R(x) the cross-interface recombination rate, K the exciton quenching rate, and $\Gamma_{\mathrm{spin}}$ the factor for spin statistics. By solving equations (2) and (4), it will be possible to establish the electron and hole density profile and the recombination density probability across the device structure. For mobility, the most accepted model was initially proposed by S. W. Liu et al. based on the concept of a bulk-heterojunction layer\cite{liu2007charge} They verified that the electron/hole mobility in the host: guest system shows a power dependence on the concentration of the individual materials and consider, based on the experimental results, that the electrical carrier in the matrix can be given by:
\begin{equation}
\begin{cases}
\mu_{p\_mix} = \mu_{p\_HTL}^{C_{HM}} \times \mu_{p\_ETL}^{C_{EM}} \\
\mu_{n\_mix} = \mu_{n\_HTL}^{C_{HM}} \times \mu_{n\_ETL}^{C_{EM}}
\end{cases}
\label{eq:mobility_mix}
\end{equation}

where C$_{HM}$+C$_{EM}$ = 1 (C$_{HM}$ material concentration for hole mobility; C$_{EM}$ material concentration for electron mobility). Also, for HOMO and LUMO levels, it is widely considered that, in the first approximation, we can assume that the HOMO level will be the highest HOMO level of all materials in the host: guest EML, whereas the LUMO level will be the lowest LUMO present. It must be noted that the eventual effect of energy levels acting as traps for the electrical carriers was not considered in the simulation. Although traps are present, we consider that the effect should be more or less similar for all devices studied. The common parameters used in the simulations are anode (ITO) work function: 4.7 eV; cathode (LiF / Al) work function: 3.0 eV; EML HOMO level: 6.0 eV; EML LUMO level: 2.9 eV; EML electron mobility: $2\times10^{-6}\ \mathrm{cm^2\,V^{-1}\,s^{-1}}$; EML hole mobility: $2\times10^{-3}\ \mathrm{cm^2\,V^{-1}\,s^{-1}}$; Average exciton diffusion length: 20 nm; Average exciton lifetime: 10 ns; Relative dielectric constant: 3.0. These parameters are allowed to change only within an acceptable physical frame, although some of them (work functions, HOMO / LUMO levels, and relative dielectric constant) was fixed (according to the device energy diagram and usual physical values) while the electrical carrier mobilities result from estimations as explained. Exciton data used the usual values found for organic semiconductors.

CBP was chosen as the host material due to the higher hole conducting capability to improve the charge balance and electron-hole pair recombination\cite{jesuraj2018recombination}. The T$_{1}$ of CBP is 2.6 eV, the hole mobility is 10$^{-3}$ cm$^{2}$ V$^{-1}$ s$^{-1}$, and the electron mobility is $2\times10^{-3}$ cm$^{2}$ V$^{-1}$ s$^{-1}$\cite{kang2007low}. The hole mobility of NPB is $1.63\times10^{-5}$ cm$^{2}$ V$^{-1}$ s$^{-1}$\cite{mangalore2019hole, chu2007hole}. TPBi was used as ETL, and the electron mobility is $3.3-8\times10^{-5}$ cm$^{2}$ V$^{-1}$ s$^{-1}$, and the T$_{1}$ level is 2.73 eV\cite{rhee2014effect, valouch2012solution}. The T$_{1}$ of TXO-TPA is 2.47 eV\cite{wang2014novel}.  The emissive layer showed PLQY of 88\% for the CBP:10 wt.\% TXO-TPA. For all the devices, the simulation of charge carrier dynamics based on the charge transport models was carried out considering current density (\textit{J}) at 40 mA cm$^{-2}$. The simulated recombination probability is summarized in Table 2 and compared with $\eta_{EQE}$.

\begin{figure*}[htpb]
    \centering
    \includegraphics[width=\textwidth]{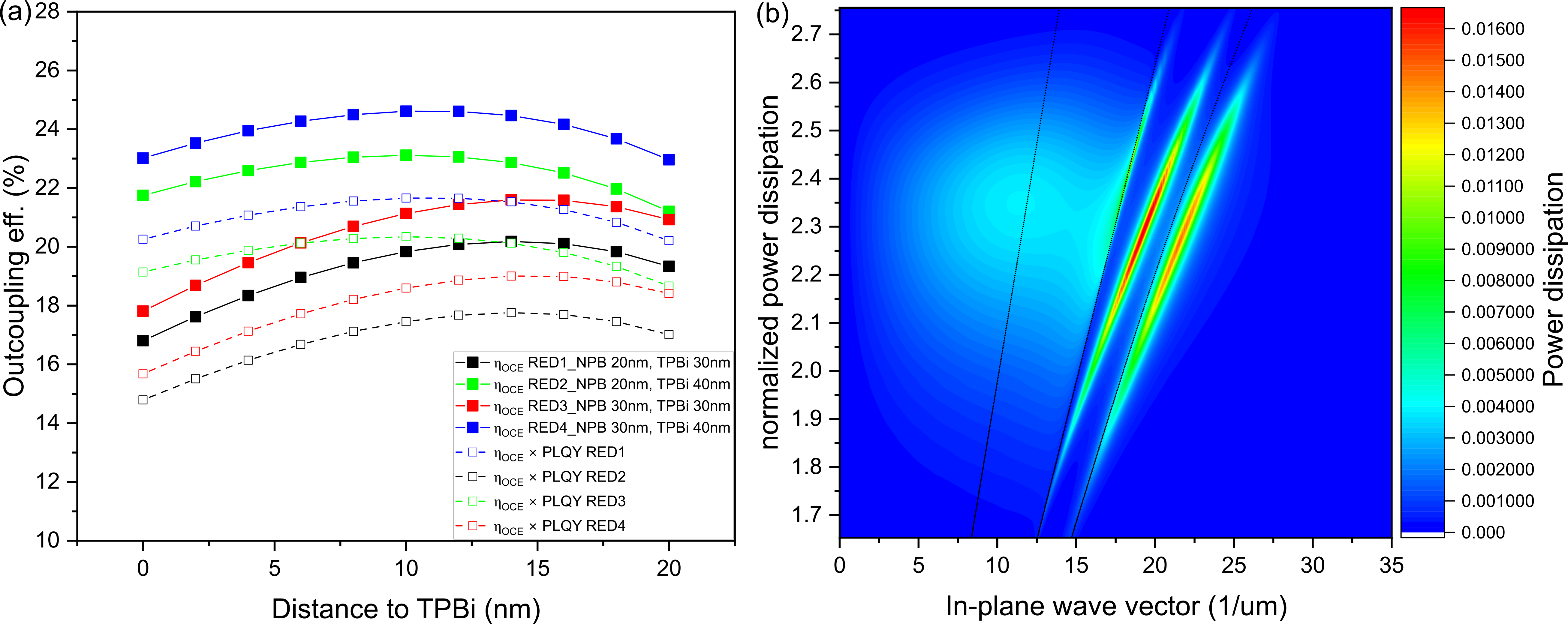}
    \caption{(a) Calculated outcoupling efficiency $\eta_{OCE}$ with emissive layer thickness distance to TPBi for all the devices (closed symbols), maximum theoretical EQE (dashed line, open symbols) represents the product of the $\eta_{OCE}$ and the PLQY of 10wt.\%TXO-TPA in CBP, (PLQY=0.88). (b) Power-dissipation plot for the RED4 device, where the outcoupled light is substantially increased compared to those RED1-3 devices, where the recombination zone is slightly near the ETL/EML interface, which results in higher losses and less photon out-coupling. (power dissipation plots for RED1-3 are shown in SI Figure~\ref{sfig1}.}
    \label{fig2}
    \end{figure*}

\subsection{Optical Simulation}\label{sec2.2}
Optical effects such as internal reflection, absorption, and surface plasmon polariton (SPP) modes can substantially restrict device performance. To incorporate these effects, we performed optical simulations following the methodology of a previous study by Leo et al.\cite{furno2012efficiency}, who used a numerical approach combining the point dipole model for radiative molecular transitions with the transfer matrix method to calculate the electromagnetic field in planar OLED structures and quantify various energy loss mechanisms within the device\cite{}. In this model, the radiative decay of excitons is described as classical electric dipole antennas emitting electromagnetic power. The outcoupling efficiency of the device is calculated based on two quantities: the total radiated power F($\lambda$) for the organic electroluminescent devices at the emitter position:
\[
F = \int S_{el}(\lambda)F(\lambda) d\lambda \\
\]
and the power component $U(\lambda)$ outcoupled at $\lambda$ calculated by integration of the spectrum of externally radiated power $K_{out}(\lambda,u)$ per unit normalized in-plane wave vector u:
\[
U = \int_0^{u_{crit}(\lambda)} K_{out}(\lambda,u) du^2 \\
\]
As:
\begin{equation}
\begin{cases}
\eta_{out} = \int S_{el}(\lambda)U(\lambda)/F(\lambda) d\lambda \\
\end{cases}
\label{eq:mobility_mix}
\end{equation}
Here, $S_{el}(\lambda)$ is the  normalized luminescence spectrum and, $U_{crit}(\lambda)$ is the maximal value of the 
in-plane wave vector allowing for far-field propagation of the emitted photons given by Snell's law:
\[
u_{crit}(\lambda) = \frac{n_0(\lambda)}{n_e(\lambda)} 
\]
where $n_0(\lambda)$ is the refractive index of the medium from which light is emitted and $n_e(\lambda)$ is the far-field medium where light is propagated into.\\

Figure~\ref{fig2} shows the outcoupling efficiency $\eta_{OCE}$ with distance to the ETL layer and power dissipation characteristics for various OLED devices. Figure~\ref{fig2}a compares the measured $\eta_{out}$ of four devices (RED1-4) with different NPB and TPBi layer thicknesses (closed symbols). It also depicts the theoretical maximum external quantum efficiency (EQE) as the product of device $\eta_{out}$ and the PLQY of 10 wt.\% TXO-TPA in CBP (open symbols, PLQY=0.88). The observed experimental values closely approach the theoretical maximum, suggesting effective optimization of device architecture and efficient exciton utilization. Notably, the RED4 device (30 nm NPB and 40 nm TPBi) demonstrates the highest outcoupling efficiency $\eta_{out}$  of 0.246 among all configurations. A simple theoretical estimate of the maximum expected $\eta_{EQE}$ can be made for practical applications as: 
\[
\eta_{EQE} = \eta_{out} \times \eta_{int}
\]
Here, the internal efficiency ($\eta_{int}$) can be approximated as: 
\[
\eta_{int} = \gamma \times \eta_{ST} \times \Phi_{PL}
\]
where $\gamma$ is the charge-balance factor (assumed to be 1 for electrically balanced devices), $\Phi_{PL}$ is the PLQY ($\Phi_{PL} = 0.88$ for the CBP: 10 wt.\% TXO-TPA), and $\eta_{ST}$ is the exciton creation efficiency ($\eta_{ST} = 1$ for TADF devices). For the outcoupling efficiency ($\eta_{out}$), we used the simulated $\eta_{out}$ value of 0.246 (Figure 2a). Thus, the maximum theoretical $\eta_{EQE}$ for our device was approximately 22\%. Our efficiency is primarily limited by the outcoupling factor (25\%), as no external extraction techniques were applied. This is consistent with other top TADF OLEDs, which often require microlenses or scattering films to push EQE beyond 30-40\%.

Figure~\ref{fig2}(b) shows the power-dissipation plot for the RED4 device (power dissipation plots for all devices are presented in Supplementary Information). The results clearly demonstrate that optimized positioning of the emissive layer significantly reduces the coupling of photons into non-radiative waveguide and surface plasmon polariton (SPP) modes. This reduction in intrinsic losses occurs because the emissive zone is effectively distanced from interfaces prone to plasmonic quenching. Consequently, a much higher fraction of generated photons are efficiently coupled into the air, leading to enhanced emission intensity and improved external quantum efficiency. These findings underline that careful design of layer thickness and positioning is critical in maximizing OLED performance by effectively suppressing internal optical losses.

\subsection{OLED preformance}\label{sec2.3}
\begin{figure*}[htpb]
    \centering
    \includegraphics[width=\textwidth]{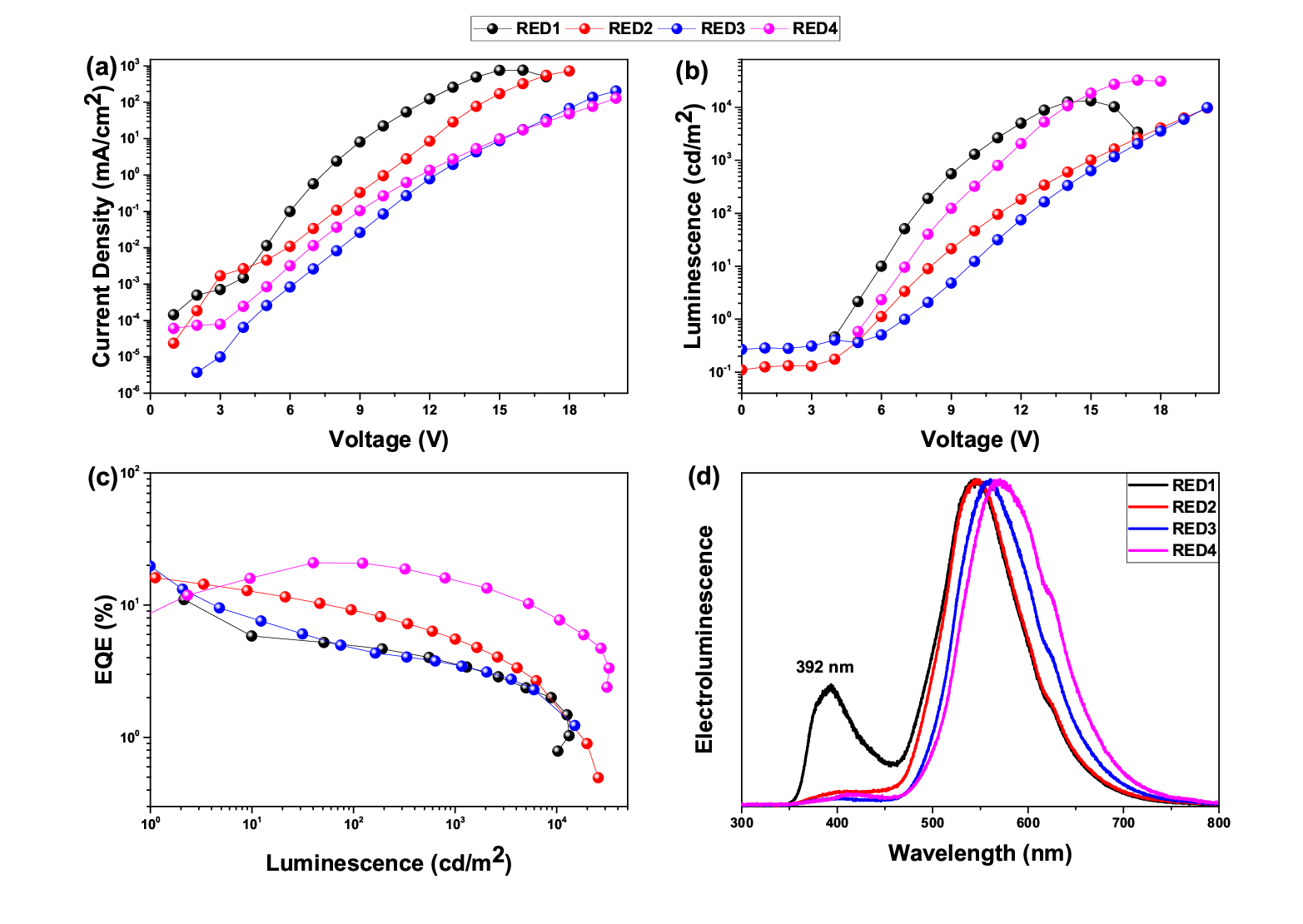}
    \caption{Device characteristics for all devices: (a) Current density-voltage, (b) luminance-voltage characteristics, (c) EQE as a function of luminance, and (d) normalized electroluminescence spectra for all the devices measured at 8V. In Figure 3d, the RED1 device emission at 392 can be clearly seen as originating from TPBi, indicating recombination happening near the ETL/EML interface. A dominant peak around 547-570 originated from TXO-TPA; the shift in emission peaks corresponds to the shifting of the recombination zone; a shoulder peak around 625 was also observed, corresponding to TXO-TPA. (CIE (x, y) coordinates for all the devices are presented in the supplementary Figure~\ref{sfig2}). The emission color of the device changed from green emission with CIE of (0.35, 0.50) in RED1 and (0.37, 0.54) in RED2 to red-orange with CIE (0.46, 0.50) in RED4.}
    \label{fig3}
\end{figure*}
To achieve an electrically balanced device employing only three organic layers, a critical analysis of the effect of HTL and ETL layer thickness on the device characteristics was established for the same host-guest EML. At the outset, four sets of devices were fabricated: RED1, RED2, RED3, and RED4, based on HTL, EML, and ETL thickness. The description of the devices is summarized in Table 1.  Figure~\ref{fig3} shows the current density vs. voltage, luminescence vs. voltage, EQE vs. luminance, and electroluminescence characteristics for all the devices.

For the RED1 device, the structure is ITO/MoO$_{3}$ (5 nm)/NPB (20 nm)/CBP-co-: TXO-TPA (10wt.\%) (20 nm)/TPBi (30 nm)/LiF (1 nm)/Al (100 nm). This configuration results in lower $\eta_{EQE}$ performance. An electrically unbalanced device seems to be obtained, primarily due to the thinner HTL/ETL thickness, resulting in unbalanced carrier mobilities. The device exhibited a maximum brightness (L$_{max}$) of 13292 cd m$^{-2}$. However, the $\eta_{CE}$ is 10 cd A$^{-1}$, $\eta_{PE}$ is 23.5 lm w$^{-1}$, and the $\eta_{EQE}$ is only 11\%. In RED1 device, the recombination occurs almost near the ETL-EML interface embedded inside ETL (Figure~\ref{fig4}b schematic). Due to this, an emission from TPBi at 392 nm can be clearly seen, and TXO-TPA at 545 nm was observed (Figure~\ref{fig3}d), resulting in a green emission with Commission Internationale de L'Eclairage (CIE) coordinates of (0.35, 0.50). Also, a shoulder peak at 625 nm was observed, similar to those previously reported emissions from TXO-TPA. These emissions can arise from different energy levels of the TADF molecule, showing weak confinement in the recombination region, perhaps due to the different electrical profiles of EML. 
 
In RED2 device with configuration of ITO/ MoO$_{3}$ (5 nm)/NPB (20 nm)/CBP:TXO-TPA (10wt.\%) (20 nm)/TPBi (40 nm)/LiF (1 nm)/Al (100 nm), due to the thicker ETL (40 nm TPBi), a better charge balance occurred, with further improvement in the recombination process (Figure~\ref{fig4}b schematic). The $\eta_{CE}$ is 24.4 cd A$^{-1}$, $\eta_{PE}$ is 28.26 lm w$^{-1}$, and the $\eta_{EQE}$ is 16.09\% with L$_{max}$ of 19888 cd m$^{-2}$. The EL emission at 545 nm was observed, while the emission from TPBi was suppressed, indicating the recombination zone shift towards EML. The CIE coordinates are (0.37, 0.54).

In the RED3 device, when HTL (NPB) is increased, a performance elevation can be seen. The device structure is ITO/ MoO$_{3}$ (5 nm)/NPB (30 nm)/CBP:TXO-TPA (10wt.\%) (20 nm)/TPBi (30 nm)/LiF (1 nm)/Al (100 nm). The $\eta_{CE}$ is 25.5 cd A$^{-1}$, $\eta_{PE}$ is 22.94 lm w$^{-1}$, and the $\eta_{EQE}$ is 19.71\% with L$_{max}$ of 15079 cd m$^{-2}$, with CIE coordinates of (0.43, 0.52), mainly due to improved carrier injection and electrical profile. It showed EL at 560 nm; this shift in the E.L. peaks usually reveals a different local interaction between host-guest and a possible solvatochromism effect, as described earlier\cite{etherington2019persistent}. The redshift in the emission spectra also could be due to the changes in exciton dynamics recombination\cite{rhee2016effects, rhee2017correlation} and can be partly due to the optical cavity formed by the device, as the emission zone position changes, the cavity interference can favor a slightly red-shifted output.

The RED4 device with the structure of ITO/ MoO$_{3}$ (5 nm)/NPB (30 nm)/CBP:TXO-TPA (10wt.\%) (20 nm)/TPBi (40 nm)/LiF (1 nm)/Al (100 nm), exhibited good luminance with high $\eta_{EQE}$ in a well-balanced device. The $\eta_{CE}$ is 37.5 cd A$^{-1}$, and the $\eta_{EQE}$ is 20.86\% with L$_{max}$ of 32798 cd m$^{-2}$. A recent high-efficiency red TADF OLED used a sophisticated multi-layer architecture to reach 40\% EQE\cite{pu2025sulfur}. In contrast, our device reaches 20.86\% with only three organic layers, highlighting a clear advantage in simplicity. Our RED4 device (37.5 cd A$^{-1}$, 14.7 lm W$^{-1}$) nearly matches the 30 cd A$^{-1}$, 30 lm W$^{-1}$ record set by a state-of-the-art red TADF emitter\cite{chen2018red}, despite our simpler design. The remaining gap in power efficacy is largely due to outcoupling losses common to all top-performing devices. In thicker 30 nm HTL and 40 nm ETL, a better charge balance is optimized inside EML, which leads to a balanced structure and optimum current density (Figure~\ref{fig4}b schematic). The device showed only one dominant EL spectrum at 570 nm corresponding to red-orange TADF TXO-TPA. In terms of performance, it is clear that the thicker n-type and p-type layers form a well-controlled charge carrier transport to facilitate the recombination zone inside the EML. The CIE coordinates are (0.46, 0.50).

\begin{table*}[!t]%
\centering %
\caption{Summary of EL performances of the OLEDs fabricated in this work.\label{tab1}}%
\begin{tabular*}{\textwidth}{@{\extracolsep\fill}llllllll@{\extracolsep\fill}}
\toprule
OLEDs & V$_{on}$ (V) & $\eta_{CE}$ (cd A$^{-1}$)$^\dagger$ & $\eta_{EQE}$ (\%)$^\ddagger$ & $\eta_{PE}$ (lm W$^{-1}$) & Lmax (cd m$^{-2}$) & CIE1931 ($x,y$) \\
\midrule
RED1 & 4.5 & 10 & 11 & 23.5 & 13292 & (0.352, 0.502)\\
RED2 & 5 & 24.4 & 16.09 & 28.26 & 19888 & (0.377, 0.543)\\
RED3 & 5 & 25.5 & 19.71 & 22.94 & 15079 & (0.431, 0.529)\\
RED4 & 5 & 37.5 & 20.86 & 14.70 & 32798 & (0.463, 0.50)\\
\bottomrule
\end{tabular*}
 \begin{tablenotes}
\item[$^\dagger$] The operating voltage at a brightness of 1 cd m$^{-2}$
\item[$^\ddagger$] Peak external quantum efficiency.
\end{tablenotes}
\label{table1}
\end{table*}

\subsection{Discussion}\label{sec2.4}
\begin{figure*}[htpb]
    \centering
    \includegraphics[width=\textwidth]{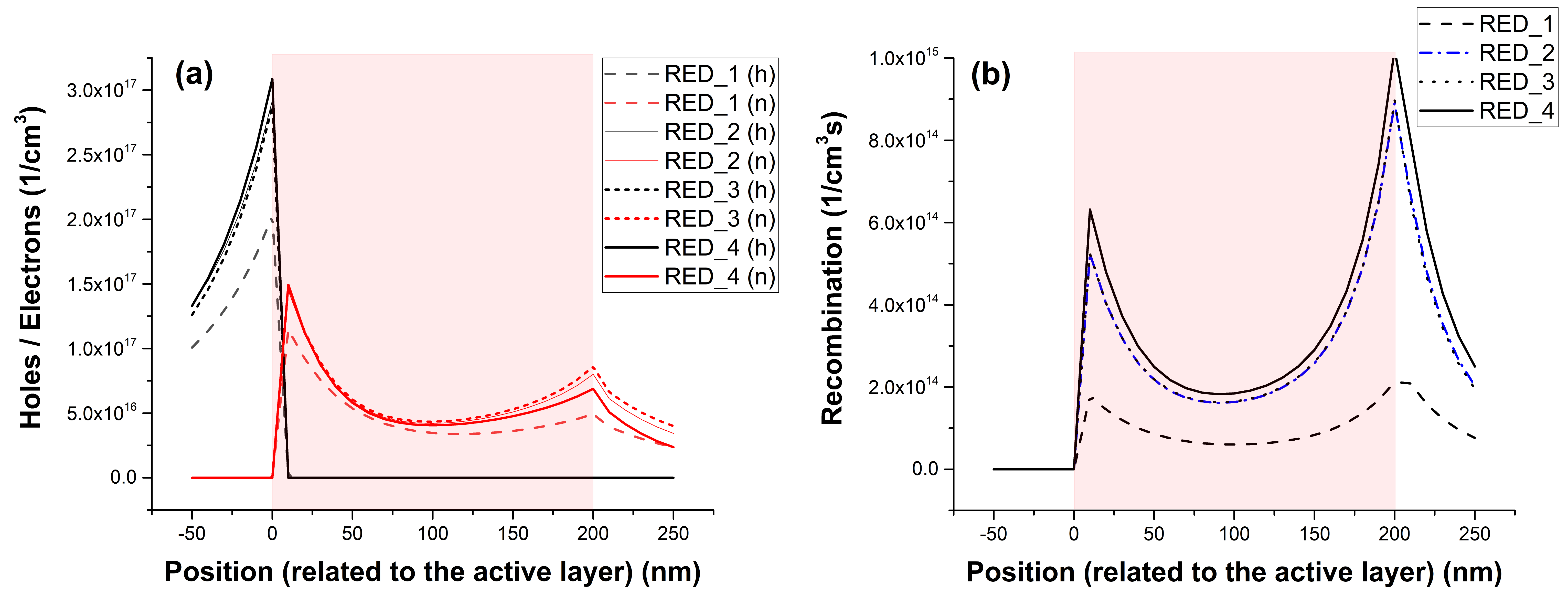}
    \caption{Electrical simulations for all the devices (Red1-4) with respect to the position inside the emissive layer: (a) hole and electrons density profile (black colour: hole density, red colour: electron density), (b) recombination zone profile inside the emissive layer, indicating the number recombination centers (cm$^{-3}$ s$^{-1}$). The shaded area represents the emission layer, starting from 0 to 200 Å. We can see that the total number of recombination centers in the RED4 device is high compared to other devices, and lowest in the RED1 device.}
    \label{fig4}
\end{figure*}
From the results, as expected, it is clear that the performance of the device depends on the architecture, where the thickness plays an important role. The HOMO levels of the compounds suggest that the energy barriers for holes and electrons are identical (same device structure). As the TADF concentration in the EML is the same in all device series, we can stipulate that charge injection through HTL/EML and ETL/EML follows the exact mechanism. Thus, the hole and electron mobilities of the organic materials have the most substantial impact on the recombination profile, density, and hence the device efficiency. The mobilities are dependent on the electrical field (driving voltage). The simulation results conclude the effectiveness of the hole and electron density profile in EML, the layer's thickness dependent, and the effect on the device efficiency. The enhanced electrical carrier profile is essential to confine the recombination zone inside the EML and not let it shift towards HTL or ETL, which is mostly affected by the thickness and electrical properties of the material\cite{rhee2017correlation}. We have simulated the recombination profile for all the devices, shown in Figure~\ref{fig4}. From the electron/hole density profiles, it is simple to observe that there are not many differences (in profiles) among all devices. However, as expected, the absolute values differ, particularly in hole injection through the HTL/EML interface. This improvement can be directly correlated with the overall performance. 

The device RED1 showed two emission peaks at 392 nm and at 545 nm, resulting in a green emission. The green emission in RED1 is not due to TXO-TPA at all, but due to incomplete exciton confinement leading to TPBi emission (fluorescent-type emitter), which leads to a lower EQE. This demonstrates that a poorly balanced device can yield a different color (here appearing green) simply because excitons form in undesired regions. By contrast, in our optimized device, the excitons recombine within the TADF-doped EML, giving the pure orange-red emission intended. The device showed a lower $\eta_{EQE}$, mainly due to the imbalanced hole and electron transport, making the recombination zone at the ETL/EML interface, resulting in a lower total recombination density of $1.83\times10^{14}$ cm$^{-3}$ s$^{-1}$. It is worth noting that increasing layer thickness slightly raised the turn-on voltage (from 4.5 V to 5.5 V), likely due to higher resistive and injection barriers. However, this is an acceptable compromise given the large gains in efficiency. Regardless of the impact of this last mechanism (injection through barriers), we can focus on pure electrical charge transport within the EML, using the usual Space Charge Limited Current (SCLC) model \ cite {pereira2012organic}. However, the loss due to radiative recombination is too high, reducing overall efficiency. When the thickness of the ETL was increased to 40 nm i.e., RED2, there was a noticeable difference in the emission spectra, diminishing the TPBi emission at 392 nm and resulting in a dominant TXO-TPA emission at 545 nm. Device performance improved drastically, with current efficiency increasing to 24.4 cd A$^{-1}$ and $\eta_{EQE}$ increasing from 11\% to 16.09\%. This is due to the much-improved electron injection from the ETL into the emissive layer, which further increases the carrier density and promotes the formation of many excitons in the EML, leading to an increased recombination density of $5.61\times10^{14}$ cm$^{-3}$ s$^{-1}$. When the thickness of NPB increased to 30 nm in device RED3, the EL spectrum shifted towards red in an ideal recombination process. No significant changes were observed in the figure of merit, suggesting better charge injection into the EML and a recombination density of $5.64\times10^{14}$ cm$^{-3}$ s$^{-1}$.  

In device RED4, increasing the ETL to 40 nm improved device performance by enhancing electron injection into the emissive layer, while maintaining recombination within the EML at a density of $8.44\times10^{14}$ cm$^{-3}$ s$^{-1}$. An excellent $\eta_{EQE}$ value of 20.86\% was achieved. The number of exciton formations increased, which is evident in the redshift of the E.L. spectra at 575 nm\cite{rhee2017correlation, rhee2016effects}. In both RED3 and RED4, we have a relatively more balanced device. However, here, the thickness of EML in all the devices is 20 nm, but the width and formation of the recombination region differ with a change in the thickness of HTL and ETL, eliminating the 392 nm parasitic emission. In all of the devices, the E.L. spectrum clearly shows only TADF emission, which means that an efficient charge transfer from the host CBP to the TXO-TPA emitter occurred. On the other hand, when the thickness of both the HTL and ETL increases, the overall series resistance should increase. This effect is observed in Figure~\ref{fig3}a and b as an increase of V$_{on}$. The RED1 exhibited higher luminance at a lower voltage because, in thin ETL and HTL layers, the carriers recombine much faster, resulting in higher luminance at a lower voltage. However, in device RED1, a higher current density is required for high luminance at a lower driving voltage. Simulation results also showed that the probability of the recombination zone increased owing to the better charge balance\cite{rhee2016effects}. This results in more exciton formation due to the participation of more molecules in the recombination, with further enhancement to the device performance\cite{song2011dependence}. It should be noted that no matter the device ETL and HTL thickness explored, there is no way to avoid some recombination loss in the ETL layer near the EML/ETL interface. Naturally, the low HOMO level difference (with no significant barrier for holes), together with some electron accumulation at such an interface, contributes to the behavior found. Increasing such a barrier (e.g., using high HOMO n-type materials such as BCP) can help to confine the recombination region in the EML; however, such a barrier will dramatically increase the overall resistance, resulting in a loss of charge injection and a further decrease of figures of merit. This can help us understand how to achieve the best trade-off we are trying to obtain. 

Our devices exhibited balanced charge injection and transport. For the outcoupling efficiency ($\eta_{out}$), we used the simulated $\eta_{out}$ value of 0.246 Figure~\ref{fig2}a). Thus, the maximum theoretical $\eta_{EQE}$ for our device was approximately 22\%, aligning with our experimental $\eta_{EQE}$ of 20.86\%. This indicates that by carefully developing the transport and EML layer thickness to enhance charge transport and exciton confinement in the EML, we achieved the maximum external efficiency possible, comparable to the best values reported in the literature for more complex device structures. Nevertheless, thickness tuning is a simple yet powerful tool, as it achieved color tuning and efficiency improvement here, which could be applied to other emitter-host combinations and further enhanced by integrating external outcoupling structures, paving the way for highly efficient, scalable OLED architectures suitable for next-generation commercial lighting and display technologies.

\subsection{Large-area OLEDs}\label{sec2.5}
\begin{figure*}[!htbp]
    \centering
    \includegraphics[width=\textwidth]{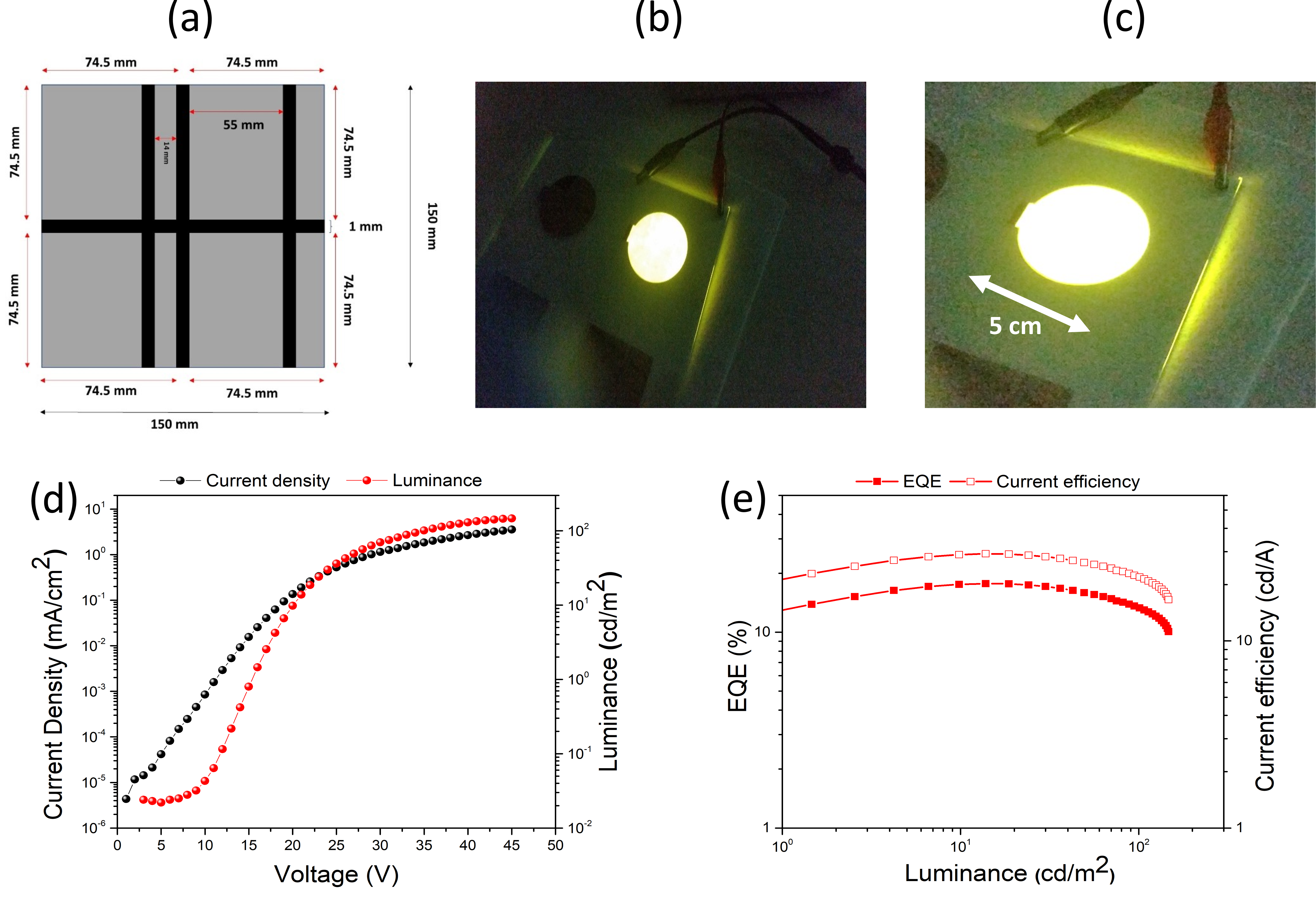}
    \caption{(a) A schematic of the large-area substrates used for the device fabrication 15$\times$15 cm$^{2}$, (b) fabricated OLED based on RED4, and (c) a pixel showing vibrant emission from the device with a diameter of 5 cm, effectively demonstrating that the device scaling is successful and the pixel emits a uniform orange light over a large area. The device characteristics of the device ITO/MoO3 (5 nm)/NPB (30 nm)/CBP:TXO-TPA (10 wt.\%) (20 nm)/TPBi (40 nm)/LiF (1 nm)/Al (100 nm):(d) current density-voltage-brightness curves, (e) $\eta_{EQE}$-current efficiency-brightness curves. The large-area device (RED4 architecture) achieves among the highest EQEs reported for large-area >10 cm$^{2}$ OLEDs via a thermal-vacuum deposition process. EQE of the large-area OLED device (20 cm$^{2}$), measured locally using a spectroradiometer with a 5 mm diameter spot. Due to measurement constraints, the shown EQE (17.7\%) likely underestimates the true total EQE, indicating excellent large-area device performance.}
    \label{fig5}
\end{figure*}

The unique nature of OLEDs allows excellent design freedom for the large production of displays and flat panels. They can be fabricated on large substrates such as glass, flexible thin plastic sheets, ceramics, metals, and fabrics. The large area OLEDs are preferred for lighting applications. Despite the advancement of the OLEDs field, the use of large-area applications is delayed, and this is mainly due to their higher production costs. However, large-area OLEDs face some problems, such as short circuits, non-uniform emission, reduced efficiency, and heat generation, which must be considered before their practical use. To address such issues, large-area fabrication was done on rigid-glass substrates via thermal evaporation. Moreover, there are not so many reports on the thermally evaporated large-area TADF OLEDs for lighting applications. The device RED4 was selected for the fabrication of the large-area application. For this, ITO-coated glass substrates with a 15$\times15$ cm$^{2}$ area were chosen to fabricate the device and assess the possibility of using the optimized structure for practical lighting applications (Figure~\ref{fig5}a-c). The device was fabricated with the structure of ITO/MoO$_{3}$ (5 nm)/NPB (30 nm)/CBP:TXO-TPA (10wt.\%) (20 nm)/TPBi (40 nm)/LiF (1 nm)/Al (100 nm).

The device exhibited excellent performance following the small-area device characteristics. For large-area devices, usually, the turn-on voltage is high at 15V. This higher turn-on voltage is ascribed to the higher series resistance in a large area and to the resistivity of the ITO-coated substrates. The device turns on at 15V, but the stability range for the device was comparatively high, and was up to 50V. All the device characteristics are shown in Figure~\ref{fig5}d,e. The device exhibited a uniform luminance, and the maximum brightness was nearly 150 cd m$^{-2}$, in accordance with practical 100 cd m$^{-2}$ applications. The device exhibited an $\eta_{EQE}$ of 17.72\% at 20V. Due to the small detection area (5 mm diameter) of our spectroradiometer, the EQE measurements reported here reflect only a local emission spot and may underestimate the true total device emission. Therefore, while the measured EQE of approximately 17.7\% for the large-area OLED (20 cm$^{2}$) is very promising, it represents a lower bound to the actual device EQE. At a practical 100 cd m$^{-2}$, the $\eta_{EQE}$ was 13.50\%, which exhibits suppressed roll-off in large-area OLEDs. Simultaneously, the $\eta_{PE}$ was 4.7 lm w$^{-1}$, and $\eta_{CE}$ was 29.21 cd A$^{-1}$. Previous large-area (>50 cm$^{2}$) OLEDs showed single-digit EQEs (<5\%)\cite{kant2023large}, whereas our work achieves 17.7\% in a 20 cm$^{2}$ red-emitting device. To our knowledge, no previous thermally evaporated TADF OLED of >10 cm$^{2}$ area has exceeded 15\% EQE; our device sets a new benchmark in this regard. Commercial OLED lighting panels typically utilize device architectures consisting of more than five organic layers, thereby increasing fabrication complexity and cost. In contrast, our simplified three-layer OLED design significantly reduces fabrication steps, providing a straightforward, cost-effective route toward scalable and practical lighting applications. For the large-area OLED, we must consider: (i) luminance non-uniformity; to achieve a useful luminous flux, brightness and the area must be considered, (ii) electrical and thermal stability, (iii) development of a simple, efficient device structure, and (iv) keeping in check the device efficiency. We must ensure that in the fabricated device, all the prerequisites were considered, and the device behaved according to all requirements for a real, practical lighting application without irreversible degradation. 

\section{Conclusions}\label{sec3}
In summary, we investigated the influence of electron and hole transport layer (ETL and HTL) thicknesses on recombination processes in simplified orange-red TADF OLEDs employing the TXO-TPA emitter in a CBP host. Our optimized three-layer OLED achieved an impressive external quantum efficiency $\eta_{EQE}$ of 20.86\% with luminance over 32800 cd m$^{-2}$. Drift-diffusion simulations demonstrated that optimizing ETL and HTL thickness significantly impacts the recombination zone, improving exciton formation and confinement in the emissive layer. By tuning transport layer thickness, $\eta_{EQE}$ improved from 11\% to 20.86\%. 

Furthermore, we successfully fabricated a large-area OLED (20 cm$^{2}$), achieving an outstanding  $\eta_{EQE}$ of 17.72\% at 20 V, comparable to the best values reported in literature for more complex structures. These findings emphasize the critical role of recombination management through transport layer engineering in simplified device architectures. This study provides valuable insights for the future development of cost-effective, high-performance OLED devices suitable for practical lighting and display applications.

\section{Experimental section}\label{sec4}
\textbf{Materials and OLED fabrication}: TXO-TPA, TPBi (2,2',2"-(1,3,5-Benzinetriyl)-tris(1-phenyl-1-H-benzimidazole)), NPB (N,N\ensuremath{{}^{\prime}}-Di(1-naphthyl)-N,N\ensuremath{{}^{\prime}}-diphenyl-(1,1\ensuremath{{}^{\prime}}-biphenyl)-4,4\ensuremath{{}^{\prime}}-diamine), and CBP were purchased from Lumtec corp. Al (99.99\%), LiF, and MoO$_{3}$ (99.99\%) were obtained from Ossila Ltd. All solvents were bought from Sigma-Aldrich. The substrates were first cleaned in an ultrasonic bath containing 2\%v/v Hellmanex solution in water, then in acetone and 2-propanol (IPA). After solvent cleaning, the substrates were treated with UV ozone for 5 minutes. Cleaned substrates were transferred to an M-Braun thermal evaporator system. All the organic layer NPB, EML, and TPBi, the LiF (1 nm) dielectric layer, and the Al (100 nm) metal layer were deposited in a vacuum at a pressure of $5\times 10^{-6}$ mbar. EML was co-evaporated with CBP and TXO-TPA. The doping concentration of TADF dopant TXO-TPA was optimized to 10 wt.\%, avoiding concentration quenching. The rate for organic materials deposition was 0.5 Å/s, 0.1 Å/s for MoO$_{3}$, LiF, and 2Å/s for Al. The active area is 0.045 cm$^{2}$ for small-area and $\approx 20$ cm$^{2}$  for large-area devices. 

\textbf{Electrical and optical characterization}: OLEDs were measured inside a glovebox using a Keithley source meter 2425 model and a Minolta LS-100 Chromameter. An Ocean Optics USB4000 spectrometer was used to measure electroluminescence spectra with a sensitivity response in the 350-950 nm wavelength range. Due to the small detection area (5 mm diameter) of our spectroradiometer, the EQE measurements reported here reflect only a local emission spot and may underestimate the true total device emission. Therefore, while the measured EQE of approximately 17.7\% for the large-area OLED (20 cm$^{2}$) is very promising, it represents a lower value to the actual device EQE.

\bmsection*{Author contributions}
M.K. and L.P. conceived the idea. M.K. fabricated the device, performed the characterization, and analyzed the data, writing the original manuscript. J.N. performed the optical simulation. J.N. and M.K. analyzed the optical simulation. M.K. and L.P. performed and analyzed the electrical stimulation. M.K. and L.P. wrote the manuscript. L.P. supervised the project. All authors contributed to the preparation of this manuscript.

\bmsection*{Acknowledgments}
Jenny and Antti Wihuri Foundation is acknowledged for financial support. Manish Kumar acknowledges funding from the Research Council of Finland through projects with decision numbers 348727 and 374566.

\bmsection*{Conflict of interest}
The authors declare no potential conflicts of interest.

\bibliography{Main}

\clearpage
\onecolumn
\setcounter{equation}{0}
\setcounter{figure}{0}
\setcounter{table}{0}
\setcounter{page}{1}
\makeatletter
\renewcommand{\theequation}{S\arabic{equation}}
\renewcommand{\thefigure}{}
\renewcommand{\@makecaption}[2]{}
\makeatother

\begin{center}
\textbf{\Large Table of Contents}
\end{center}
An efficient, large-area orange-red OLED with a simplified three-layer structure employing a third-generation thermally activated delayed fluorescence (TADF) emitter is demonstrated. This straightforward design enables high performance and scalability, making it highly suitable for general lighting applications.

\vspace{1cm}

\begin{center}
    \includegraphics[scale=3.5]{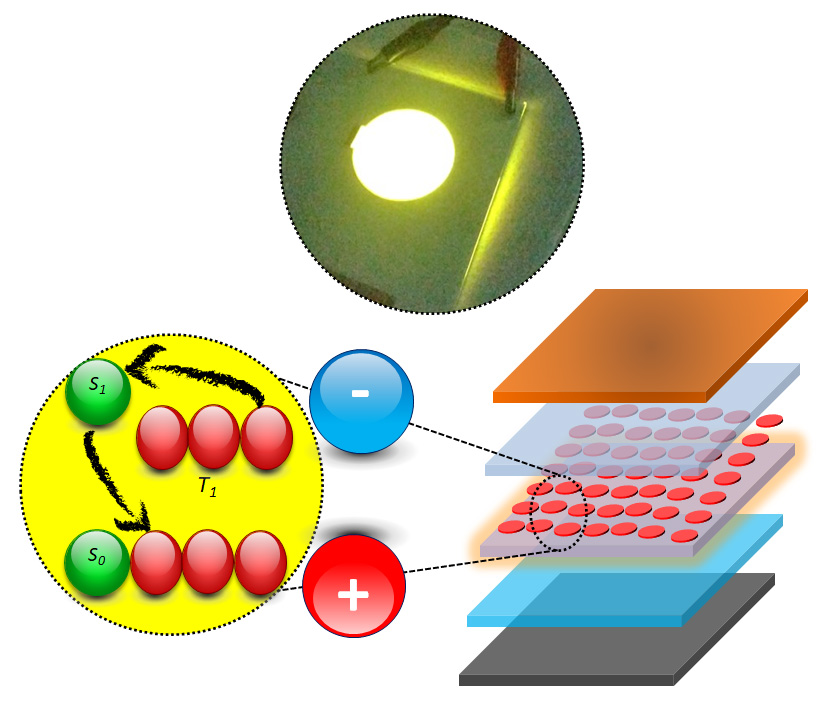}
\end{center}

\clearpage
\onecolumn
\setcounter{equation}{0}
\setcounter{figure}{0}
\setcounter{table}{0}
\setcounter{page}{1}
\makeatletter
\renewcommand{\theequation}{S\arabic{equation}}
\renewcommand{\thefigure}{S\arabic{figure}}
\renewcommand{\@makecaption}[2]{
\vskip3pt
\fontsize{9}{11}\selectfont
\noindent\textbf{#1} #2\par
\vskip4pt
}
\makeatother
\begin{center}
{\Large\bfseries Supporting Information\par}
\end{center}
\begin{center}
{\Large\bfseries Transport Layer-Driven Optimization of Color and Efficiency in a TADF OLEDs for Large-Area Applications\par}
\end{center}
\noindent
Manish Kumar$^{*}$, Jawid Nikan, and Luiz Pereira$^{*}$\\
\noindent
Corresponding authors: manish.kumar@utu.fi, luiz@ua.pt
\section*{Contents}
\begin{figure}[htpb]
\centerline{\includegraphics[width=\columnwidth]{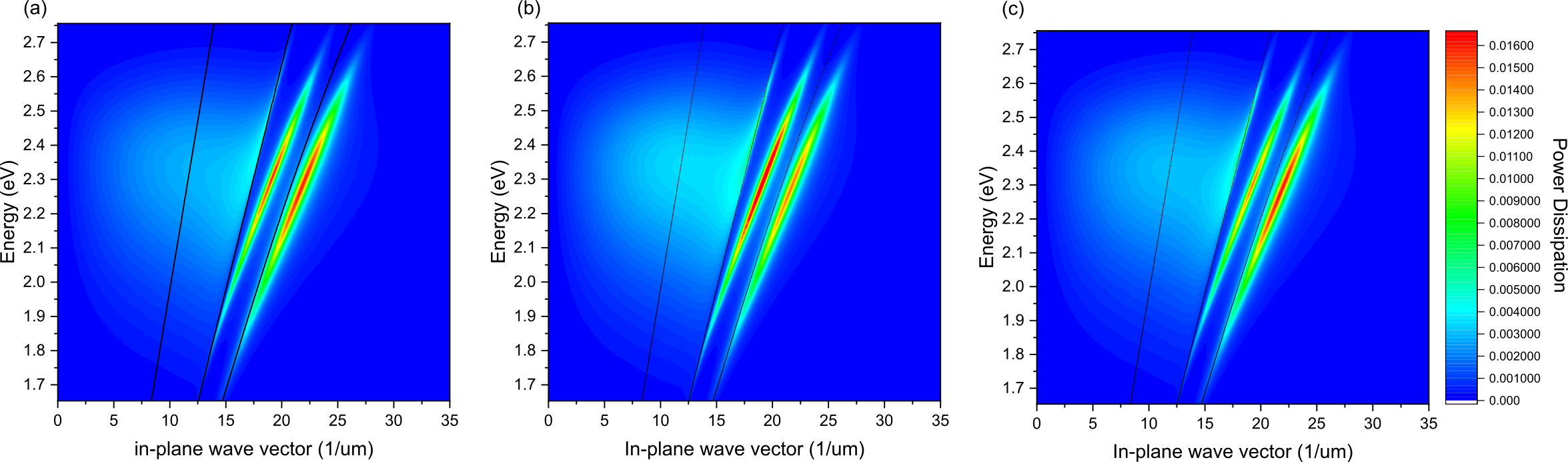}}
\caption{Power-dissipation plot for the RED1, RED2 and RED3 devices.
\label{sfig1}}
\end{figure}

\begin{figure}[htpb]
\centerline{\includegraphics[width=\columnwidth]{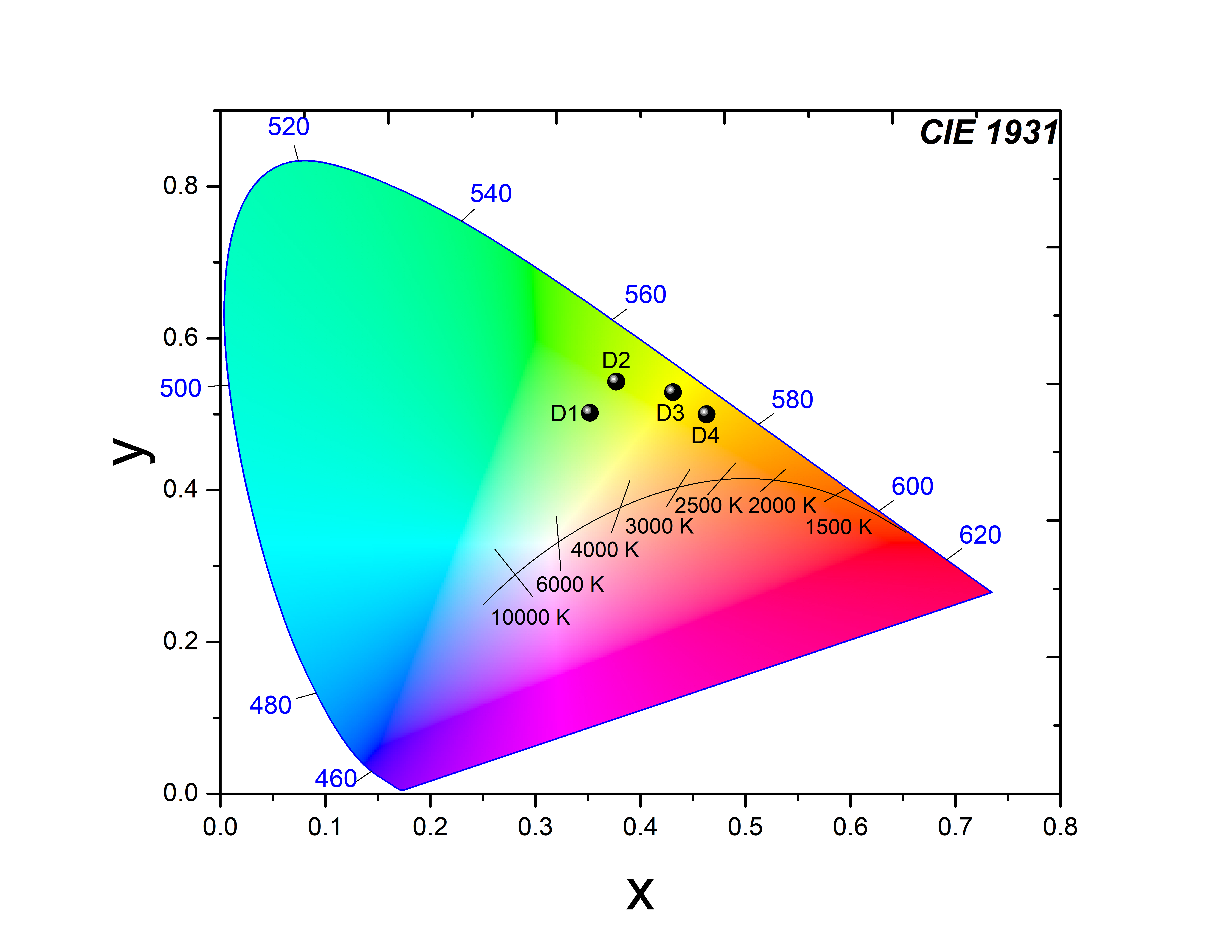}}
\caption{CIE (x,y) coordinates for RED1, RED2, RED3 and RED4 devices.
\label{sfig2}}
\end{figure}

\textbf{Supplementary Figure \ref{sfig1}}. Power-dissipation plot for the RED1, RED2, and RED3 devices.\\
\textbf{Supplementary Figure \ref{sfig2}}. CIE (x,y) coordinates for RED1, RED2, RED3 and RED4 devices.\\

\end{document}